\documentstyle[12pt,epsfig]{article}
\newfont{\sfb}{cmssbx10 scaled 1400}
\newfont{\bigsf}{cmssbx10 scaled 1600}

\begin{document}
\begin{center}
{\LARGE\bf  Is there window for ``supersoft'' Pomeron in 
$J/\psi$ photoproduction at low energy?}

\vspace{1cm}
D. Roy$^{a}$, T. Morii$^a$, H. Toki$^b$ and A.I. Titov$^{c}$\\

\vspace{0.5cm}
{\it $^{a}$Graduate school of Science and Technology and
\\ Faculty of Human development, Kobe University,
\\ Kobe 657-8501, Japan} \\

\vspace{0.5cm}
{\it $^{b}$Research Center of Nuclear Physics, Osaka University\\
Osaka 567-0047, Japan}\\

\vspace{0.5cm}
{\it $^{c}$Bogolyubov Laboratory of Theoretical Physics, JINR\\ 141980 Dubna, Russia}\\

\vspace{1.0 cm}
\end{center}
\begin {center}
{\bf Abstract}
\end{center}

\baselineskip=22pt
The low energy $J/\psi$ photoproduction cross-section has been studied 
upon the basis of the Pomeron model. To incorporate the discrepancy 
between experimental data and predictions by conventional models, i.e. 
the sum of the soft Pomeron with intercept 1.08 and the hard Pomeron 
with intercept 1.418, a Regge trajectory associated with a scalar 
meson ($f,a$) exchange which we call ``supersoft'' Pomeron, is 
introduced additionally.  To distinguish between the conventional 
model and this new additional Pomeron, observations related to other 
polarization observables in upcoming polarized experiments are discussed.

\vspace{2.0em}

PACS number(s): 13.85.L; 13.60.H; 13.85

\vfill\eject

\baselineskip 22pt
\noindent
%{\Large 1. Introduction}\\

\noindent
1. It is well-known that exclusive  photoproductions of light
vector mesons $\rho$, $\omega$, and $\phi$ have been
characterized by a weak dependence of the cross-section
on the photon-proton center of mass energy $W$ and by a diffractive peak,
i.e. small scattering angle of the vector meson with respect
to the incident photon direction.
This behavior has been explained by the Vector Dominance Model (VDM) 
and Regge theory~\cite{ZEUS98}.
Regge theory has dominated various aspects of high energy particle 
physics~\cite{1}.  It is considered to be applicable in the region 
where $W^{2}$ is much greater than other variables. But surprising 
thing is that it works sometimes very well even at considerably smaller 
energy, close to the threshold~\cite{Will98}.
As stated above, the data on production of the light vector mesons 
at high energies have been well explained in this theory by the 
exchange of a single nonperturbative 
Pomeron ~\cite{DL84-86,DL87a,DL88,Cud90,LM95,PL96}, which is known as the
``soft-Pomeron''~\cite{DL98} with trajectory 
\begin{eqnarray}
\alpha_s(t)=\alpha(0)+\alpha't=1.0808 + 0.25t.\ 
\label{s-p-trajectory}
\end{eqnarray}
When the energy dependence of the vector-meson production 
cross-section is parameterized as $W^\delta$  
(in the Regge theory approach, $\delta=4(\alpha_s(0)-1-\alpha'/b)$, 
where $b$ is the slope of the $t$-distribution with typical value 
$b=10$ GeV$^{-2}$), then for light vector mesons, 
$\delta\simeq0.22$ is found to be a good  value for reproducing the data.

One widely discussed  problem arises
when this model is applied to $J/\psi$ productions at large energy 
for $W>10$ GeV~\cite{ZEUS98}.
The cross-section for exclusive $J/\psi$ productions by quasi-real 
photons $(Q^2=0$) at HERA is observed to rise more steeply with $W$.
Parameterization shows that in this case $\delta\simeq 0.8$,  
contrary to the theoretical expectation with $\delta\simeq 0.2$.
This steep energy dependence is  different from the ``soft'' behavior 
of light vector mesons, and is known as the ``hard'' behavior 
of $J/\psi$ meson.  There are two basic approaches for
explaining this discrepancy: perturbative two-gluon 
contribution~\cite{5,6,7} and the contribution of the hard 
Pomeron by Donnachie and Landshoff~\cite{DL98}.

Another problem is related to the low energy region. 
Previous analysis by Donnachie and Landshoff (DL) for 
description of various hadronic reactions with hadrons consisting 
of light $u,d,s$ quarks shows that the low energy behavior may be 
successfully parameterized by a single effective Regge pole whose 
trajectory has intercept $\alpha=0.5475$~\cite{DL92}. For simplicity, 
in this paper we call this contribution as a ``supersoft'' trajectory 
or "supersoft" Pomeron.  Note that the physical background for introducing
this trajectory is not obvious in the case of light quarks/hadrons 
because the conventional low-order meson-exchange may contribute 
just at low energy.  However, the situation is different
in photoproduction of $J/\psi$-meson which consists of $c,\bar c$ 
quark and thus, conventional meson exchange dynamics is forbidden.
In this case the idea of contribution of the additional ``supersoft'' 
Pomeron seems to be more natural.  Besides the above mentioned 
DL-supersoft Pomeron, another possible contribution of the 
trajectory inspired by ($J^\pi=0^+, M^2\sim 3$ GeV$^2$)
glueball predicted by Lattice QCD and QCD motivated models, 
has been also discussed in literature~\cite{10,11}.
In this paper, we concentrate mainly on this subject of 
$J/\psi$ photoproduction and analyze the possible manifestation 
of the supersoft Pomeron at low energy.\\

%{\Large 2. Pomeron Models}\\
\noindent
2. The Pomeron being a gluonic assembly has long been established 
both theoretically and experimentally~\cite{14}.  Based on the 
Pomeron model, to explain the ``hard'' or steep $W^2$ behavior 
of $J/\psi$ photoproductions, many models
have been proposed so far, among which two typical models are 
remarkable: (a) the two-gluon models motivated by perturbative QCD 
widely discussed in literature~\cite{5,6,7}  and (b) the 
phenomenological model motivated by the non-perturbative QCD and Regge 
theory~\cite{DL98}, i.e. hard Pomeron exchange with intercept 1.418.
Our main interest here is the low energy region where the model 
(a) does not work, and by considering recent doubt on detection of 
BFKL-Pomeron at finite energy (present energy region)~\cite{CDL96}, 
in this paper we will use the modified Pomeron model of~\cite{TLTS99} 
with incorporation of the Pomeron trajectories from the 
Donnachie and Landshoff analysis of~\cite{DL98,DL92}, to 
illustrate the high energy $J/\psi$-photoproduction.

%%%%%%%%%%%%%%%%%%%%%%%%%%%%%%%%%%%%%%%%%%%%%%%%%%%%%%%%%%%%%%%%%%%%%%%%%%%

In the vector dominance model, an incoming photon first converts 
into a vector meson which then scatters diffractively from the nucleon.
Within QCD, a microscopic model of Pomeron was proposed by Donnachie 
and Landshoff~\cite{DL87a}, where an incoming photon first converts
into a quark-antiquark pair and then exchanges a Pomeron with the nucleon.
After the interaction, the quark and  antiquark recombine to form 
the $J/\psi$ meson~\cite{DL87a}.
Here we adopt the simplified version of the DL model to define the 
Pomeron exchange amplitudes for $J/\psi$ photoproductions with 
$Q^2=0$ which has been verified by the nonperturbative QCD 
models~\cite{DL84-86,LM95,PL96}.  In this model which reproduces 
rather well the vector meson photoproduction and diffractive 
electro-dissociation, the soft Pomeron behaves like a $C=+1$ 
isoscalar photon which has been justified by Landshoff and Natchmann 
in nonperturbative two gluon model~\cite{LN87}.
In fact, the same suggestion has been made in~\cite{DL98} for the 
hard Pomeron.  If it originates from the hard perturbative two 
gluon exchange, then this suggestion seems to be natural, as can be 
seen from the direct calculation of the corresponding loops~\cite{5}.
However, for "supersoft" Pomeron, this suggestion is not self-evident.  
Moreover, if one assumes that it originates from the scalar meson 
($f,a$) exchange~\cite{DL92}, then it is more natural to consider 
that the corresponding effective vertices are described by an 
"effective" scalar exchange.  The difference between these two 
different suggestions may be seen in polarization observables as 
we show below.

The DL model leads to the invariant amplitude of the $J/\psi$ 
photoproduction in the form
\begin{eqnarray}
T_{fi}^{P_n}=\bar{u}_{m_{f}}({p}')M_{0}^{P_n}\epsilon^\ast_{\psi\mu} 
M^{P_{n}\mu\nu}\epsilon_{\gamma\nu}u_{m_{i}}(p), 
\label{inv-ampl}
\end{eqnarray}
with
\begin{eqnarray}
M^{P_{n}\mu\nu}=F_{\alpha}^{{n}}\Gamma^{P_{n}\alpha,\mu\nu}, 
\label{inv-amp1}
\end{eqnarray}
where $n = s, h$ and $m$ for soft, hard and supersoft Pomeron 
trajectories, respectively;
$u(p)$ is the Dirac spinor; $\epsilon_{\gamma\nu}$ and 
$\epsilon_{\psi\mu}$ are the polarization vectors of the photon 
and the $J/\psi$ meson, respectively. 
$F_{\alpha}^{n}$ $(n=s, h~{\rm and}~m)$ describes the 
Pomeron-nucleon vertex: 
$F_{\alpha}^{h,s}=\gamma_\alpha,\,\, F_{\alpha}^m=1$.
$\Gamma^{P_{n}\alpha,\mu\nu}$ is related to the Pomeron-$J/\psi$ 
coupling\cite{TLTS99}:
\begin{eqnarray}
\Gamma^{P_s\alpha,\mu\nu} = \Gamma^{P_h\alpha,\mu\nu} = 
g^{\alpha\nu}\, k^\mu - k^\alpha \,g^{\mu\nu},\qquad\qquad
\Gamma^{{P_m}\mu\nu}=({ k^\mu q^\nu} - k\cdot q\, g^{\mu\nu})/M_V , 
\label{Gam2}
\end{eqnarray}
where $k$ and $q$ are the 4-momentum of the incoming photon and the 
outgoing $J/\psi$ meson, respectively, and the transversality 
conditions $M^{P_n\mu\nu}\cdot q_\mu = M^{P_n\mu\nu}\cdot k_\nu=0$ 
are fulfilled.
The factor $M_{0}^{P_n}$ in Eq.(2) is given by the conventional 
Regge pole amplitude
\begin{eqnarray}
M_{0}^{P_n}={C_V}_n\,F(t)\,F_n(s,t) e^{-i\frac{\pi}{2}\alpha_{n}(t)} 
\left(\frac{s-s_n}{s_0}\right)^{\alpha_{n}(t)}.
\label{ampl3}
\end{eqnarray}
Parameter $s_n$ is introduced to extend the standard DL-model to 
the low energy region. Practically we use the natural "threshold" 
scale for this parameter: $s_n=(M_N+M_{J/\psi})^2$.
The function $F(t)$ is an overall form factor~\cite{LM95} 
\begin{eqnarray}
F(t) = F_{V}\cdot F_{N}(t),
\end{eqnarray}
where,
\begin{eqnarray}
&&F_{N}(t) = \frac{(4M_N^2-2.8t)}{(4M_N^2-t)(1-{t}/{0.7})^2},\nonumber\\ 
&&F_{V}(t)=\frac{M_V^2}{(M_V^2-t)^2}\,\frac{\mu^2}{2\mu^2 + M_V^2-t},
\end{eqnarray}
The constant ${C_V}_n$ is given by
\begin{eqnarray}
 {C_V}_n=18\,{C_0}_n\,s_0 \,\beta^2
\left(\frac{\alpha\Gamma_{V\to e^+e^-}}{M_V}\right)^{1/2}, 
\end{eqnarray}
with $V\equiv J/\psi$, $\beta_0=4$ GeV$^{-2}$, 
$\mu^2=1.1$ GeV$^2$~\cite{DL88,LM95} and the other parameters are standard.
The correcting function $F_n$ in Eq.(\ref{ampl3}) is given by\cite{TLTS99} 
\begin{eqnarray}
F_{n}^{-2} = \frac{1}{4}\Gamma_{\mu\nu}^{P_{n}\alpha} \Gamma_{\mu'
\nu'}^{P_{n}{\alpha'}}
Tr\{F_\alpha^{n}(p\!\!\!/+M_N)F_{\alpha'}^{n}(p'\!\!\!/+M_N)\} 
(g^{\mu\mu'}-q^{\mu}q^{\mu'}/M_{J/\psi}^2)g^{\nu\nu'} /4M_N^2.
\end{eqnarray}
The corresponding Regge trajectories read 
\begin{eqnarray}
&&\alpha_h(t)=1.418 + 0.1t,\,\,\,\, {\rm for~hard~Pomeron},\nonumber\\ 
&&\alpha_s(t)=1.0808 + 0.25t,\,\,\,\, {\rm for~soft~Pomeron},\nonumber\\ 
&&\alpha_m(t)=0.5475+ 0.25t,\,\,\, \,{\rm for~supersoft~Pomeron}.
%\nonumber\\
%\alpha_g(t)=-0.75 + 0.25t,
\end{eqnarray}
The constant factors ${C_{0}}_n$ are chosen to reproduce 
d$\sigma/dt|_{t=0}$ from threshold and up to $W$=100 GeV: 
${C_0}_s\simeq 0.58$,
${C_0}_h\simeq 0.05 \, {C_0}_s$,
${C_0}_m\simeq 0.33 \, {C_0}_s$.
The result of our fit is shown in 
%Fig.~\ref{f2}, 
Fig. 1, where we display differential cross section 
$d\sigma/dt$ as  function of $W$ at $t=t_{\rm max}$ (or $J/\psi$ 
production angle $\theta=0$).
Experimental data are taken from Refs.~\cite{ZEUS98,experiment}.  
The right panel shows the calculation at all available energy region.
The left panel shows only the low energy region.
One can see that the difference between calculations with soft and 
hard Pomerons alone
and data at low energy is about factor of 2. Inclusion of the 
supersoft trajectory improves the fitting to the data.  
In principle, the same effect can be obtained from incorporation of 
the supersoft trajectory $\alpha_g(t)=-0.75 + 0.25t$ inspired by glue
ball dynamics~\cite{11}, though in this case we have to put the 
threshold parameter $s_n=0$ and correspondingly change ${C_0}_n$ in Eq.(8).

Another manifestation of the  ``supersoft'' Pomeron may appear 
in the spin-density matrix elements of the $J/\psi$ decay, 
as described in the following.\\

%{\large 3-2. Density matrix elements of the $J/\psi$ meson}\\ \noindent
\noindent
3. The angular distribution of $J/\psi\rightarrow a + b$ is defined 
by~\cite{SSW}
\begin{eqnarray}
\frac{dN}{d\cos\Theta d\Phi}=
\sum|T_{\lambda_{f},\lambda_{\psi};\lambda_{i}, \lambda_{\gamma}}\cdot 
M_{\lambda_{\psi}}(\Theta,\Phi)|^2, 
\end{eqnarray}
where $\Theta$ and $\Phi$ are the decay and azimuthal angles of $a$ or $b$
in the rest system of the $J/\psi$ meson, respectively. For convenience, 
we use the Gottfried-Jackson (GJ) frame.
$M_{\lambda_{\psi}}$ is the decay amplitude of the $J/\psi$ meson 
with helicity $\lambda_{\psi}$ and given by, 
\begin{eqnarray}
M_{\lambda}(\Theta,\Phi)=C\sqrt{\frac{3}{4\pi}} 
D_{\lambda\lambda_{ab}}^{1*}(\Theta,\Phi,-\Theta), 
\end{eqnarray}
where $\lambda_{ab}=\lambda_{a} -\lambda_{b}$ is the helicity 
difference between $a$ and $b$. The constant $|C|^2$ is 
proportional to the decay width and if we are working with 
normalized angular distributions, it drops out of the final result, 
and hence we can set $C$ = 1.
Using Eq.(12), one can express the normalized distribution in the 
following form,
\begin{eqnarray}
\frac{dN}{d\cos\Theta d\Phi}=W(\cos\Theta,\Phi) =
\frac{3}{4\pi}\sum_{\lambda\lambda_{ab}} 
D_{\lambda\lambda_{ab}}^{1*}(\Theta,\Phi,-\Theta) \rho_{\lambda\lambda'}
D_{\lambda'\lambda_{ab}}^{1} (\Theta,\Phi,-\Theta)
\end{eqnarray}
where $\rho_{\lambda\lambda'}$ is the $J/\psi$ spin density matrix 
element given by
\begin{eqnarray}
\rho_{\lambda\lambda'}=
\frac{1}{N}\sum_{\lambda_f,\lambda_{\gamma}; \lambda_{i}\lambda_{\gamma}'}
T_{\lambda_{f},\lambda;\lambda_{i}, \lambda_{\gamma}}
\rho(\gamma)_{\lambda_{\gamma}\lambda_{\gamma}'} 
T_{\lambda_{f},\lambda';\lambda_{i},\lambda_{\gamma'}}, 
\end{eqnarray}
with the normalization factor,
\begin{eqnarray}
N = \sum |T_{\lambda_{f},\lambda;\lambda_{i},\lambda_{\gamma}}|^2, 
\end{eqnarray}
and $\rho(\gamma)_{\lambda_{\gamma}\lambda_{\gamma}'}$ is the 
incoming photon density matrix.
The angular distributions due to unpolarized incident photons 
are determined by $\rho_{\lambda,\lambda'}^0$ with 
$\rho(\gamma)_{\lambda_{\gamma}\lambda_{\gamma'}} = 
\delta_{\lambda_{\gamma}\lambda_{\gamma'}} $.
For circularly (linearly) polarized
incident photons, the angular distributions are calculated from 
$\rho^0$ and $\rho^3$($\rho^0,\rho^1$ and $\rho^2$).
For illustrating the possible manifestation of the "supersoft" 
trajectory, we are limited to $\rho^0$ matrix elements.
The whole idea is to show how these matrix elements behave 
depending on the models proposed (for example, on the supersoft 
Pomeron model).
The results for three such matrix elements are shown in 
%Fig.~\ref{f3} 
Fig. 2 as function of the $J/\psi$ production angle 
in c.m.s. at $E_\gamma=10$ GeV.
The three models shown here  are
(i) the spin-conserving model,
(ii) the sum of hard and soft Pomerons, 
(iii) the sum of hard, soft and supersoft trajectories.
The case of pure scalar exchange for supersoft trajectory 
corresponds to the spin-conserving model, i.e. the model (i).
The large difference between (ii) and (iii) suggests that 
this study can actually be a strong test for the existence 
of the supersoft Pomeron.\\

%{\Large 4.Conclusion and discussion}\\

In summary, we have analyzed the possible manifestation of 
the supersoft Pomeron inspired by the scalar meson $f,\,a$ 
(or glueball) exchange dynamics at low energy. We show that 
inclusion of this trajectory considerably improves agreement 
with experimental data on unpolarized cross section.
Definite prediction for the spin-density matrix element has been done.
However, we emphasize that the present investigation is very 
exploratory, owing to the lack of the precise data at low energy. 
It is hightly desirable to obtain more data from the new facilities 
such as LEPS of SPring-8 in Japan and TJNAL. The polarization 
observables are most useful for future theoretical investigation.

\vfill\eject

\begin {center}
{\Large \bf Figure captions}
\end{center}

\begin{description}
\item[Fig.\ 1:] The differential cross section $d\sigma/dt$ of 
$J/\psi$ photoproduction at $t=t_{\rm max}$:
(a) total available energy region,
(b) low energy region.
Notation: $P_s,\,P_h,\,P_{m}$ show the separate contributions of 
the soft, hard and supersoft trajectories, respectively; $P_s+P_h$ 
is for the coherent sum of the soft and hard Pomerons, $\Sigma$ is 
for the coherent sum of the all Pomeron contributions.
\item[Fig.\ 2:] Spin-density matrix elements $\rho^0_{00}$,
$\rho^0_{01}$  and $\rho^0_{11}$ for left, middle and left panels, 
respectively, as function of the $J/\psi$ production angle 
in c.m.s. at $E_\gamma=10$ GeV.
Notations: SS represents the prediction for the pure supersoft 
trajectory or the spin-conserving model, S+H is for the sum of 
hard and soft Pomerons, $\Sigma$ is for the total 3-Pomeron model.
\end{description}

\newpage
\begin{figure}
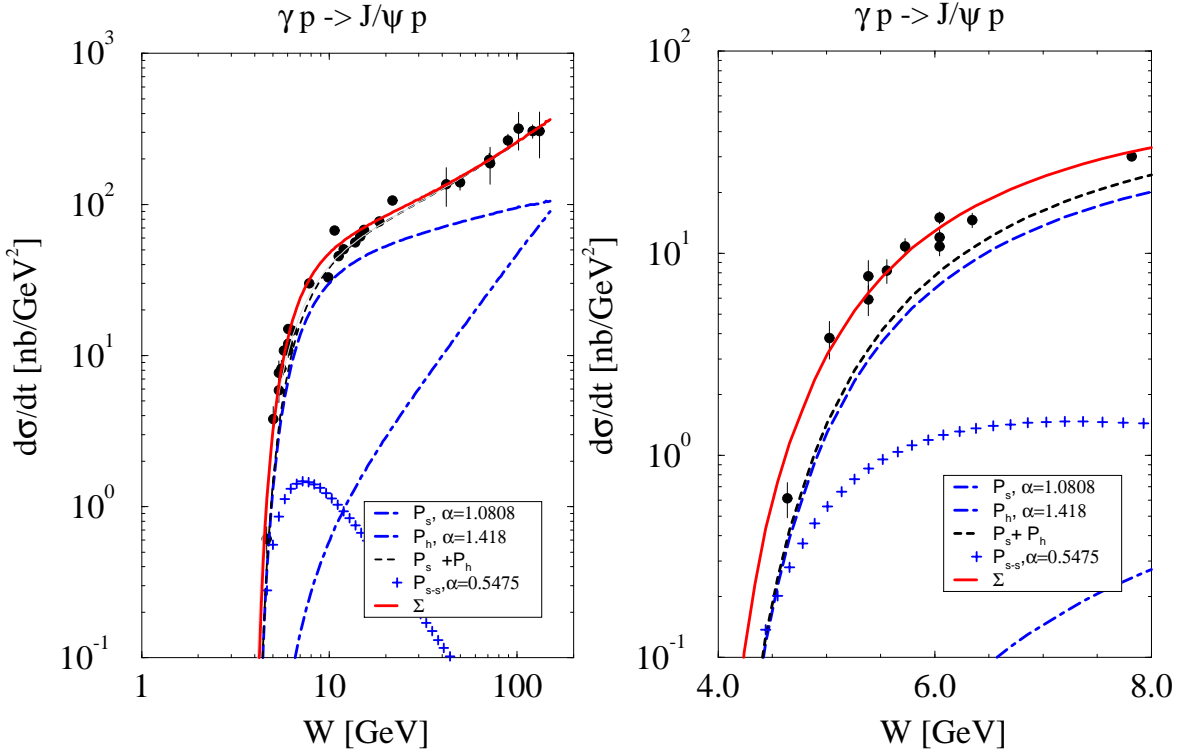

\vspace{1cm}
\centering{
\epsfig{file=f2l.epsi, width=0.435\columnwidth} 
\epsfig{file=f2r.epsi, width=0.45\columnwidth}}
\caption{The differential cross section $d\sigma/dt$ of 
$J/\psi$ photoproduction at $t=t_{\rm max}$:
(a) total available energy region,
(b) low energy region.
Notation: $P_s,\,P_h,\,P_{m}$ show the separate contributions 
of the soft, hard and supersoft trajectories, respectively; $P_s+P_h$ 
is for the coherent sum of the soft and hard Pomerons, $\Sigma$ is 
for the coherent sum of the all Pomeron contributions.}
\label{f2}
\end{figure}

\begin{figure}
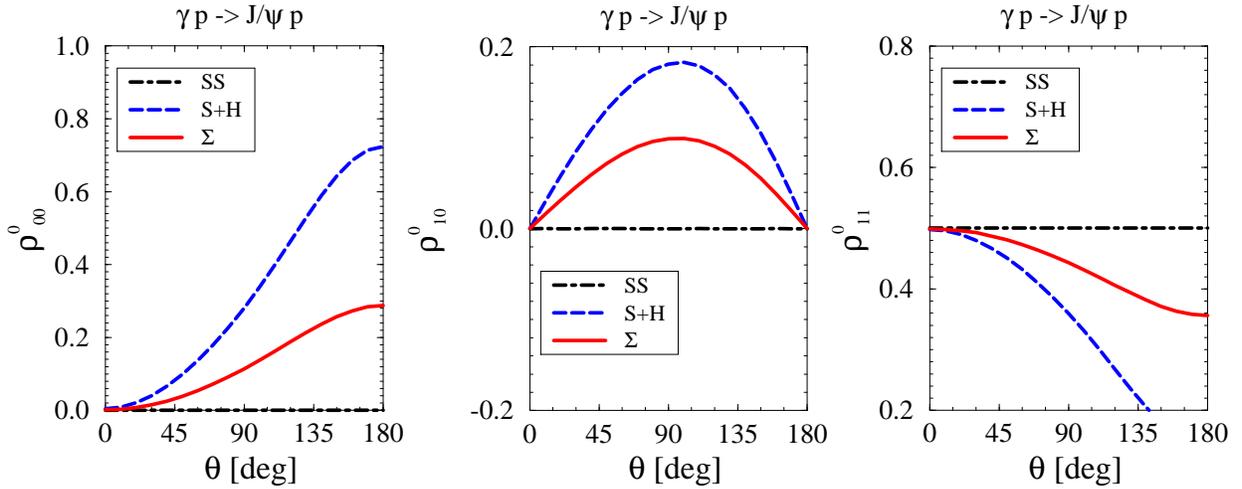

\vspace{1cm}
\centering{
\epsfig{file=f3.epsi, width=0.3\columnwidth} 
\epsfig{file=f4.epsi, width=0.318\columnwidth} 
\epsfig{file=f5.epsi, width=0.3\columnwidth}} 
\vspace{0.5cm}
\caption{Spin-density matrix elements $\rho^0_{00}$,
$\rho^0_{01}$  and $\rho^0_{11}$ for left, middle and left panels, 
respectively, as function of the $J/\psi$ production angle 
in c.m.s. at $E_\gamma=10$ GeV.
Notations: SS represents the prediction for the pure supersoft 
trajectory or the spin-conserving model, S+H is for the sum of hard 
and soft Pomerons, $\Sigma$ is for the total 3-Pomeron model.} 
\label{f3}
\end{figure}

\end{document}